\documentclass[10pt,twocolumn]{article}
\usepackage{times,color,graphicx}
\usepackage{amsmath,amssymb}
\usepackage{biophysics}

\title{Profile Conditional Random Fields for Modeling Protein Families with Structural Information}
\author{Akira R. Kinjo\\
(email: akinjo@protein.osaka-u.ac.jp)\\
Institute for Protein Research, Osaka University\\
Suita, Osaka, 565-0871, Japan}
\begin{document}
\maketitle
\begin{abstract}
A statistical model of protein families, called profile conditional random fields (CRFs), is proposed. This model may be regarded as an integration of the profile hidden Markov model (HMM) and the Finkelstein-Reva (FR) theory of protein folding. While the model structure of the profile CRF is almost identical to the profile HMM, it can incorporate arbitrary correlations in the sequences to be aligned to the model. In addition, like in the FR theory, the profile CRF can incorporate long-range pairwise interactions between model states via mean-field-like approximations. We give the detailed formulation of the model, self-consistent approximations for treating long-range interactions, and algorithms for computing partition functions and marginal probabilities. We also outline the methods for the global optimization of model parameters as well as a Bayesian framework for parameter learning and selection of optimal alignments.
\end{abstract}

\section*{Introduction}
Protein sequence alignment is one of the most fundamental techniques in 
biological research. Since the early methods have been 
proposed\cite{NeedlemanANDWunsch1970,SmithANDWaterman1981,Gotoh1982}, 
techniques for protein sequence alignment have made a huge progress
 toward the detection of very weak homology\cite{DurbinETAL,ProteinBioinfo}.
Today, most advanced methods incorporate some kind of information obtained 
from multiple sequence alignments in terms of sequence profiles\cite{GribskovETAL1987} or 
position-specific scoring matrices (PSSM). In sequence profiles, 
such as used in PSI-BLAST\cite{AltschulETAL1997},
scores for amino acid substitutions are made to be position-specific so that
subtle evolutionary signals can be embedded in each 
site\cite{KinjoANDNakamura2008}.
This in turn makes homology search more sensitive.
Profile hidden Markov models (HMM)\cite{KroghETAL1994,DurbinETAL} further 
elaborate the sequence profile methods so that deletions and insertions are 
also made position-specific. Although powerful, these methods do have some 
limitations. The profile methods (including profile HMMs) assume that each 
position in a profile is independent of other positions which makes it 
difficult to incorporate long-range correlation among different sites. The 
importance of long-range correlations is evident when one takes into account 
the tertiary structure of a protein in which residues far apart along the 
sequence are in contact to define the specific native structure. In practice,
one can supplement a plain sequence profile with some structural information
as in three-dimensional (3D) profile\cite{BowieETAL1991} or 
threading\cite{JonesETAL1992}, but such combined approaches remain inherently 
\emph{ad hoc}. In case of profile HMMs, it is extremely difficult, if not 
impossible, to employ such an approach since the inclusion of site-site 
correlations, both short-range and long-range, may break the probabilistic 
framework of the model.

In order to incorporate long-range correlations into an HMM-like model in a 
well-defined manner, we present in this paper the theoretical formulation of 
a model based on conditional random fields (CRF)\cite{CRF}. 
Various CRF-based models have been successfully applied to many problems
in biological domains including pairwise protein sequence 
alignment\cite{DoETAL2006}, gene prediction\cite{DeCaprioETAL2007}, and 
protein conformation sampling\cite{ZhaoETAL2008}, to name a few.
CRFs share many of the advantages of HMMs while being able to handle 
site-site correlations. In the context of profile CRFs, 
we need to distinguish two types of site-site correlations. 
One is the correlations within the sequence 
which is to be aligned to a CRF model; the other is those among the sites 
within the model. The profile CRF model proposed in this paper has 
no limitations for incorporating the both types of correlations, 
although some approximations are necessary for the latter type in 
practical applications. Without the model sites correlations, the profile CRF 
model may be regarded as a generalization of the profile HMM. Unlike profile 
HMMs, profile CRFs can incorporate many kinds features of the sequence
in terms of feature functions.
With the model sites correlations, the profile CRF model may be regarded as a 
generalization of the self-consistent molecular field theory of Finkelstein
and Reva\cite{FinkelsteinANDReva1991,FinkelsteinANDReva1996a,FinkelsteinANDReva1996b}, which, in turn, is a generalization of the Ising model in one-dimension 
(1D). 

In this paper, we first present the model structure of the profile CRF, 
provide some examples of possible feature functions, and derive 
some approximations for treating long-range correlations between model sites.
Next, we present algorithms for computing partition functions, marginal 
probabilities, and optimal alignments, followed by methods for parameter 
learning based on multiple sequence alignments.
Since our purpose here is to present the formulation and algorithms,  actual 
implementation of the method and experimentation thereof are left for 
future studies. Nevertheless, we believe that the method presented here
should serve as a firm basis for the analysis of protein sequences and 
structures.
\section*{Theory}
\subsection*{Profile conditional random field model}
\begin{figure}[tb]
  \centering
  \includegraphics[width=7cm]{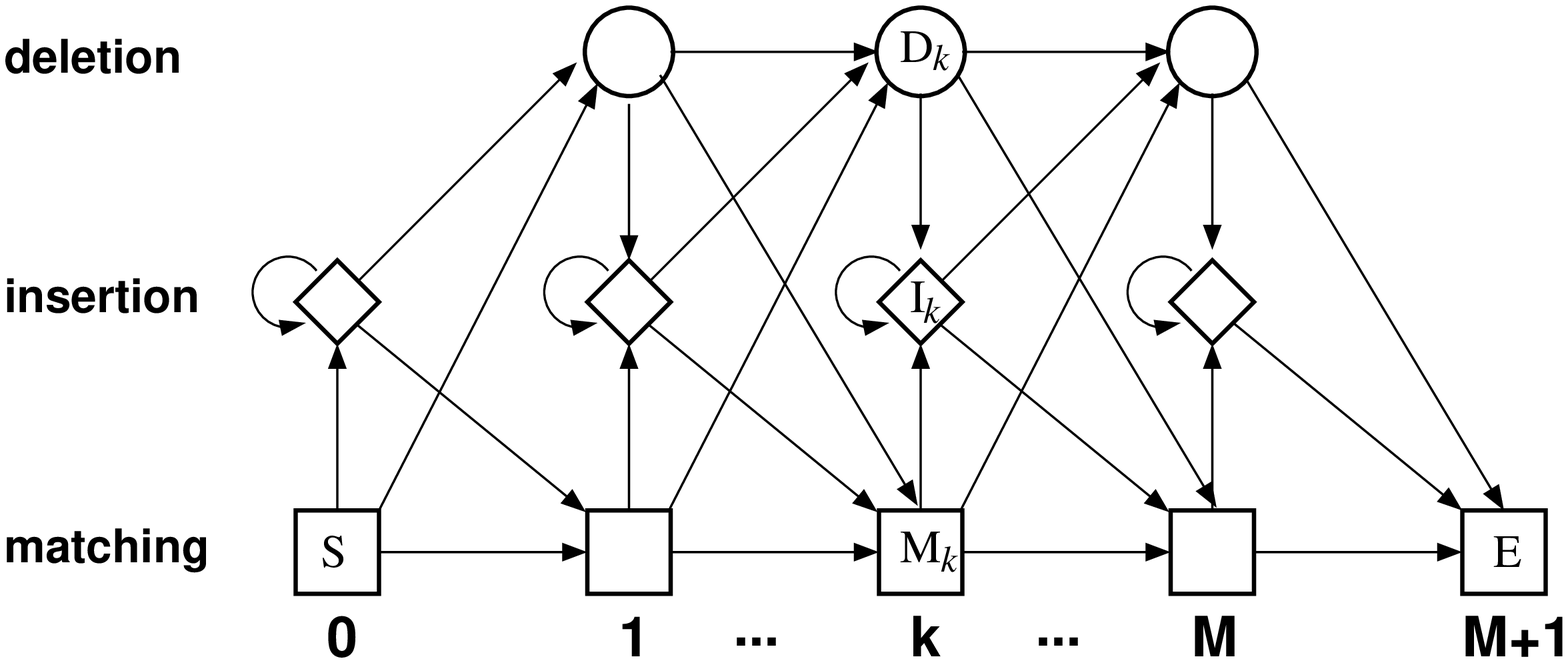}
  \caption{The model structure of a profile conditional random field (CRF).
Squares, diamonds, and circles are matching, insertion, and deletion states, respectively. The start and end states are labeled with ``S'' and ``E'' in the squares.}
  \label{fig:model}
\end{figure}
We model a protein family (or a multiple sequence alignment) in an analogous 
manner as profile HMMs\cite{DurbinETAL,KroghETAL1994} (Fig. \ref{fig:model}). 
A profile CRF model $\mathcal{M}$ is formally defined as a tuple of 
four components:
\begin{equation}
  \label{eq:def}
  \mathcal{M} = (M,\mathcal{S}, \mathcal{F}, \theta)
\end{equation}
where $M$ is the \emph{length} of the model $\mathcal{M}$, 
$\mathcal{S}= \{M_k, I_k, D_k\}$ is a set of \emph{states}
indexed by \emph{model sites} $k = 0, 1, \cdots, M, M+1$. For each 
site $k$ ($1 \leq k \leq M$), there are a \emph{matching state} $M_k$, 
an \emph{insertion state} $I_k$, and a \emph{deletion state} $D_k$. 
For $k = 0$, there are only one matching state $M_0$ and one insertion 
state $I_0$; for $k = M+1$, there is only one matching state $M_{M+1}$.
The matching states at the termini $M_0$ and $M_{M+1}$ are also called 
\emph{start state} and \emph{end state}, respectively, for the reason that 
will be apparent in the following. The model sites with $k = 1, \cdots, M$ 
may be regarded as the core sites of the protein family.
The third component, $\mathcal{F}$, is a set of \emph{feature functions} 
which are associated with
model states ($\mathcal{S}$). Each feature function maps an amino acid
sequence and its site indexes to a real number depending on model sites.
The last component, $\theta$, is 
a set of parameters or \emph{external fields}, each of which is associated with
a feature function in $\mathcal{F}$. Together with feature functions, the external fields are used for evaluating alignments between the model and 
amino acid sequences.
The details of these terms will be clarified below. 
In a profile CRF model, the feature functions must be given \emph{a priori} 
and the values of external fields are learned from a multiple sequence 
alignment (MSA).  

The objective is to align a protein sequence  
$\mathbf{x} = x_1x_2\cdots{}x_L$ (called \emph{target sequence}) to the model. 
An alignment between a target sequence $\mathbf{x}$ and 
a CRF model is an ordered sequence of pairs of target sites and 
model states (called site-state pairs in the following):
\begin{equation}
  \label{eq:ali}
  \mathcal{A} = \{ (0,M_0), (1,y_1), \cdots, (i, y_i), \cdots, (L+1,M_{M+1})\}
\end{equation}
where $y_i = S_k \in \{M_k, I_k, D_k\}_{k = 0, \cdots, M+1}$. 
The pair $(i,y_i)$ reads as ``the target site $i$ is matched to the model 
state $y_i$.''
It is assumed that 
if $i \leq j$ and $y_i = S_k$ and $y_j = S_l$, then $k \leq l$.
An alignment always starts at the start state and ends at the end state so that
the pairs $(0,M_0)$ and $(L+1,M_{M+1})$ are fixed in any alignments.
In an alignment, not all transitions from one site-state pair to another
are possible. Allowed transitions are listed in Table \ref{tab:transition} and
depicted in Fig. \ref{fig:model} by arrows.
\begin{table}[tb]
  \centering
  \caption{Allowed transitions of site-state pairs. $i$ and $k$ indicate a site of a target sequence and a site of a CRF model, respectively.}
  \begin{tabular}{lll}\hline
    $(i,M_k)$ & $\to$ & $(i+1, M_{k+1})$\\
    $(i,M_k)$ & $\to$ & $(i+1, I_{k})$\\
    $(i,M_k)$ & $\to$ & $(i, D_{k+1})$\\
    $(i,I_k)$ & $\to$ & $(i+1, M_{k+1})$\\
    $(i,I_k)$ & $\to$ & $(i+1, I_{k})$\\
    $(i,I_k)$ & $\to$ & $(i, D_{k+1})$\\
    $(i,D_k)$ & $\to$ & $(i+1, M_{k+1})$\\
    $(i,D_k)$ & $\to$ & $(i+1, I_{k})$\\
    $(i,D_k)$ & $\to$ & $(i, D_{k+1})$\\\hline
  \end{tabular}
  \label{tab:transition}
\end{table}
By convention, the match to a delete state $(i,D_k)$ means that the deletion 
resides between the $i$-th and $(i+1)$-th positions of the sequence.
For example, an alignment of an 8-residue target sequence to an $M=7$ profile
 CRF model might be given as 
\begin{multline}
  \mathcal{A} = (\mathbf{x}, \mathbf{y}) =\\
  \label{eq:ali2}
  \{ (0, M_0), (1,I_0), (2,M_1), (3, M_2), (3,D_3),(4,M_4), \\
(5,M_5), (6,I_5), (7,I_5), (8,M_6), (8,D_7), (9, M_8)\}.
\end{multline}
As can be inferred from this example, $(i,M_k)$ indicates that the $i$-th 
residue matches the $k$-th core site of the model, $(i,I_k)$ indicates that
there is an insertion at the $i$-th site in the target sequence, and 
$(i,D_k)$ indicates that there is a deletion between $i$-th and $(i+1)$-th
sites of the target sequence. In terms of ordinary sequence alignment, the alignment in Eq. (\ref{eq:ali2}) may be expressed as 
\begin{equation}
  \begin{array}{cccccccccc}
    - & M_1 & M_2 & M_3 & M_4 & M_5 & - & - & M_6 & M_7 \\
    x_1 & x_2 & x_3 & - & x_4 & x_5 & x_6 & x_7 & x_8 & - \\
  \end{array}
\end{equation}
where the `$-$' signs in the upper and lower rows indicate insertions (corresponding to $I_k$) and deletions ($D_k$) in the model sites.

Alignments are evaluated in terms of a set of feature functions 
$\mathcal{F} = \{s_{S}^{\alpha}, t_{S,S'}^{\beta}, u_{S,S'}^{\gamma}\}$.
Three types of feature functions are distinguished, namely,
 singlet feature functions $s_{S}^{\alpha}(\mathbf{x},i)$, 
doublet feature functions $t_{S,S'}^{\beta}(\mathbf{x},i)$, 
and pairwise feature functions $u_{S,S'}(\mathbf{x},i,j)$. 
The singlet feature function (SFF)
$s_{S}^{\alpha}(\mathbf{x},i)$ is a real-valued function representing some 
feature $\alpha$ of the target sequence when $y_i = S$; 
the doublet feature function (DFF)
$t_{S,S'}^{\beta}(\mathbf{x},i)$ is also a real-valued function representing 
some feature $\beta$ when $y_{i^{-}} = S$ 
and $y_{i} = S'$. Here, $i^{-}$ is the \emph{predecessor} of $i$ defined as
\begin{equation}
  \label{eq:predecessor}
  i^{-} = \left\{
    \begin{array}{ll}
      i & \mbox{(if $y_i = D_k$),} \\
      i-1 & \mbox{(if $y_i = M_k$ or $I_k$). }
    \end{array}
\right.
\end{equation}
In general, $\alpha$ may depend on $S$ and $\beta$ may 
depend on $S$ and $S'$.
The singlet and doublet feature functions are called local or short-ranged
since the former represents interactions at one model site and the latter,
 interactions between two adjacent model sites for which transitions are 
allowed.
The pairwise feature function (PFF) $u_{S,S'}^{\gamma}(\mathbf{x},i,j)$, 
representing some feature $\gamma$, is defined for $y_i = S$ and $y_j = S'$. 
While singlet and doublet feature 
functions are local, pairwise feature functions are non-local in the 
sense that $S$ and $S'$ can be any pair of the model states, 
not necessarily those for which direct transitions are allowed.

Each of singlet, doublet or pairwise feature functions 
is coupled with a parameter called 
an external field: $\lambda_{S}^{\alpha}$ for $s_{S}^{\alpha}$,
$\mu_{S,S'}^{\beta}$ for $t_{S,S'}^{\beta}$, and $\nu_{S,S'}^{\gamma}$ for 
$u_{S,S'}^{\gamma}$. That is, $\theta = \{\lambda_{S}^{\alpha}, \mu_{S,S'}^{\beta},
\nu_{S,S'}^{\gamma}\}$. The product of a feature function and its coupled 
external field yields the score of the corresponding feature 
when a particular target site is aligned to a model state.
For example, the product $\lambda_{S}^{\alpha}s_{S}^{\alpha}(\mathbf{x},i)$ 
is the score of the feature $\alpha$ when the target site $i$ is aligned to 
the model site $S$. In the formulation of CRF, it is convenient to employ 
an analogy to statistical physics.
Thus, the negative total score of an alignment is interpreted as the total 
energy, and the normalization factor for the conditional probability of 
alignments as the partition function of the target sequence.

Given an alignment between the model and the sequence, 
the total energy of an alignment $\mathcal{A} = (\mathbf{x},\mathbf{y}) = 
\{(0,M_0), \cdots, (i, y_i), \cdots, (L+1,M_{M+1})\}$ 
is defined by
\begin{multline}
  \label{eq:energy}
  E(\mathbf{y},\mathbf{x},\theta) = \\ 
-\sum_{\{i\}}\left[
\sum_{\alpha}\lambda_{y_i}^{\alpha}s_{y_i}^{\alpha}(\mathbf{x},i)
+ \sum_{\beta}\mu_{y_{i^{-}},y_{i}}^{\beta}t_{y_{i^{-}},y_{i}}^{\beta}(\mathbf{x}, i)
\right]\\
- \sum_{\{i<j\}}\sum_{\gamma}\nu_{y_i,y_j}^{\gamma}u_{y_i,y_j}^{\gamma}(\mathbf{x},i,j)
\end{multline}
where the summation over $\{i\}$ means summing along the alignment 
$(\mathbf{x},\mathbf{y})$ (there can be multiple occurrences of the same index 
$i$ due to the matching to deletion states); 
the double summation for $i<j$ is also similarly defined.
The partition function of this system is thus given by
\begin{equation}
  \label{eq:pfunc}
  Z(\mathbf{x},\theta) 
   =  \sum_{\{\mathbf{y}\}}\exp[-E(\mathbf{y},\mathbf{x},\theta)/T]
\end{equation}
where the summation is over all possible alignments, and $T$ is 
the temperature (in energy unit).
The conditional probability of obtaining a particular alignment 
$\mathcal{A} = (\mathbf{x},\mathbf{y})$ for a given $\mathbf{x}$ is 
\begin{equation}
  \label{eq:probali}
  P(\mathbf{y}|\mathbf{x},\theta) = \frac{\exp[-E(\mathbf{y},\mathbf{x},\theta)/T]}{Z(\mathbf{x},\theta)}
\end{equation}
which is also called the likelihood of the alignment in the following. The log-likelihood is defined by 
\begin{eqnarray}
  \label{eq:ll}
  L(\theta|\mathbf{x},\mathbf{y}) & = & \log P(\mathbf{y}|\mathbf{x},\theta) \nonumber\\
& = & -E(\mathbf{y},\mathbf{x},\theta)/T - \log Z(\mathbf{x},\theta).
\end{eqnarray}
From here on, we assume $T = 1$ unless otherwise stated.
The derivatives of the log-likelihood with respect to the parameters,
$\partial L/\partial \theta$, are 
useful both for parameter learning and for deriving approximations.
For singlet terms, they are given as
\begin{equation}
  \label{eq:dlambda}
  \frac{\partial L(\theta|\mathbf{x},\mathbf{y})}{\partial \lambda_{S}^{\alpha}} = 
\sum_{\{i\}}
s_{S}^{\alpha}(\mathbf{x},i)
[\delta_{S,y_i} - P(S|\mathbf{x},i)]
\end{equation}
where $\delta_{S,y_i}$ is Kronecker's delta and $P(S|\mathbf{x},i)$ is the marginal probability that $i$-th site of the target sequence is aligned to 
the state $S$ of the model, i.e., $y_i = S$.
Similarly for the doublet terms,
\begin{multline}
  \label{eq:dmu}
  \frac{\partial L(\theta|\mathbf{x},\mathbf{y})}{\partial \mu_{S,S'}^{\beta}} = 
\\
\sum_{\{i\}}t_{S,S'}^{\beta}(\mathbf{x},i)
[\delta_{S,y_{i^{-}}}\delta_{S',y_{i}} - P(S, S'|\mathbf{x},i)]
  \end{multline}
where $P(S, S'|\mathbf{x},i)$ is the marginal probability that 
$y_{i^-} = S$ and $y_i = S'$.
Finally for the pairwise terms,
\begin{multline}
  \label{eq:dnu}
  \frac{\partial L(\theta|\mathbf{x},\mathbf{y})}{\partial \nu_{S,S'}^{\gamma}} =
\\  \sum_{\{i<j\}}u_{S,S'}^{\gamma}(\mathbf{x},i,j)
[\delta_{S,y_i}\delta_{S',y_j} - P(S, S' | \mathbf{x}, i, j)]
\end{multline}
where $P(S,S' | \mathbf{x},i,j)$ is the marginal probability that $y_i = S$ and 
$y_j = S'$.
Either when parameters are optimal for a given alignment or 
when the alignment is optimal for given parameters, 
we have $\partial L / \partial \theta = 0$.

\subsection*{Feature functions}
Although we are focused on the formulation of the profile CRF model, it is 
instructive to provide some concrete examples for feature functions. 
It should be stressed, however, that the actual selection of feature functions 
will require careful experimentation to maximize the effectiveness of the 
profile CRF framework.
\paragraph{Singlet feature functions.}
Singlet feature functions represent compatibility measures between a model 
state and a target sequence. It may depend on the whole target sequence 
as well as on single amino acid residues. One simple SFF may be such that
\begin{equation}
  \label{eq:singlet1}
  s_{M_{k}}^{R}(\mathbf{x},i) = \delta_{x_i,R}
\end{equation}
where $R$ is one of the 20 standard amino acid residue types.
It is implicitly assumed that this feature function is defined only
when $y_i = M_k$. The same assumption applies throughout the following 
discussion.

If the target sequence is accompanied by its PSSM, the above SFF 
(Eq. \ref{eq:singlet1}) can be generalized as
\begin{equation}
  \label{eq:singlet2}
  s_{M_{k}}^{\mathrm{PSSM}(R)}(\mathbf{x},i) = \mathrm{PSSM}(i,R)
\end{equation}
where $\mathrm{PSSM}(i,R)$ is the value of the PSSM for residue type $R$ at 
site $i$.

SFFs can also depend on multiple sites of the target sequence. For example,
let us partition amino acid residues into either hydrophobic (1) or 
hydrophilic (0). Let $b_7(\mathbf{x},i)$ be a binary word 
encoding\cite{KawabataANDDoi1997} function 
of the 7-residue sub-sequence $x_{i-3}\cdots x_{i+3}$. Then, the SFF
\begin{equation}
  \label{eq:singlet3}
  s_{I_k}^{0000000}(\mathbf{x},i) = \delta_{0000000,b_7(\mathbf{x},i)}
\end{equation}
may enhance insertions at highly hydrophilic regions of the target sequence.
Similarly, the SFF
\begin{equation}
  \label{eq:singlet4}
  s_{I_k}^{0011011}(\mathbf{x},i) = \delta_{0011011,b_7(\mathbf{x},i)}
\end{equation}
may enhance the matching at $\alpha$-helical regions since the binary pattern
0011011 is typical in $\alpha$ helices. There are $2^7 = 128$ types of binary 
words for 7-residue segments, and we can incorporate all of these in a single 
profile CRF model.

If either predicted or observed structural information is available for 
the target sequence, we may define, for example,
\begin{equation}
  \label{eq:singlet5}
  s_{M_k}^{H}(\mathbf{x},i) = \delta_{H,SS(i)}
\end{equation}
where $SS(i)$ indicates the secondary structure of site $i$.
\paragraph{Doublet feature functions.}
Doublet feature functions represent the feasibility of transitions from one 
site-state pair to another. One trivial example is those that do not depend
on the target sequence at all. For example, the DFF
\begin{equation}
  \label{eq:doublet1}
  t_{M_{k},I_{k}}^{-}(\mathbf{x},i) = 1
\end{equation}
may be regarded as a feature representing a gap (insertion) opening.
Similar sequence-independent DFFs can be defined for all the allowed 
state transitions.

Of course, DFFs can be made to be target sequence-dependent. Take the binary 
word encoding example again. For example, the following DFF
\begin{equation}
  \label{eq:doublet2}
  t_{M_k,D_{k+1}}^{001101}(\mathbf{x},i) = \delta_{001101,b_6(\mathbf{x},i)}
\end{equation}
may help to suppress deletions at $\alpha$-helical regions, since the 
pattern 001101 is typical in $\alpha$-helices in which deletions are less 
likely to occur.

\paragraph{Pairwise feature functions.}
With pairwise feature functions, it is possible to incorporate some kind of
correlations between two states that are not directly connected by transitions.
Such correlations are most easily grasped in the context of the 
tertiary structure of a protein. Suppose that there is a known structure in 
a protein family to be modeled as a profile CRF, and that structure contains
 a pair of contacting residues which correspond to the matching states 
$M_k$ and $M_l$. We may define
\begin{equation}
  \label{eq:pair1}
  u_{M_k,M_l}^{\mathit{contact}(R,R')}(\mathbf{x},i,j) = \delta_{x_i,R}\delta_{x_j,R'}
\end{equation}
where $R$ and $R'$ are amino acid residue types. We can define different 
PFFs for different kinds of interactions such as hydrogen bonds, 
salt bridges, hydrophobic contacts, etc. Also, sequence-dependence may be 
made more complex. We can combine contacts with binary word encoding, 
for example.

\subsection*{Approximations for pairwise interactions}
If there are no pairwise terms, exact partition functions
and exact optimal alignments for profile CRF models can be computed 
efficiently by dynamic programming just as in profile HMMs.
With pairwise terms present, however, the computation of exact solutions 
is intractable. In order to 
make computations feasible, we need to make some approximations. 
More specifically, we will derive a Bethe approximation, 
which is further simplified to a mean-field approximation.

Observe, first, that the pairwise terms can be rearranged as
\begin{multline}
  \label{eq:longapprox}
  \sum_{\{i<j\}}\sum_{\gamma}\nu_{y_i,y_j}^{\gamma}u_{y_i,y_j}^{\gamma}(\mathbf{x}, i, j)
= \\
\sum_{\{i<j\}}\sum_{\gamma}\sum_{S,S'}\nu_{S,S'}^{\gamma}u_{S,S'}^{\gamma}(\mathbf{x}, i, j)\delta_{S,y_i}\delta_{S',y_j}.
\end{multline}
When the alignment is optimal, we have ${\partial L(\theta|\mathbf{x},\mathbf{y})}/{\partial \nu_{S,S'}^{\gamma}} = 0$ (Eq. \ref{eq:dnu}), 
hence the following:
\begin{multline}
  \sum_{\{i<j\}}u_{S,S'}^{\gamma}(\mathbf{x},i,j)\delta_{S,y_i}\delta_{S',y_j} =\\
 \sum_{\{i<j\}}u_{S,S'}^{\gamma}(\mathbf{x},i,j)P(S, S' | \mathbf{x}, i, j).
\end{multline}
Using this relation, the pairwise terms are arranged as
\begin{multline}
 \sum_{\{i<j\}}\sum_{\gamma,S,S'}\nu_{S,S'}^{\gamma}u_{S,S'}^{\gamma}(\mathbf{x}, i, j)
\delta_{S,y_i}\delta_{S',y_j}
 = \\
\sum_{\{i\}}\sum_{\gamma,S}\tilde{u}_{S}^{\gamma}(\mathbf{x},i)P(S | \mathbf{x},i)
  \label{eq:bethe1}
\end{multline}
where $\tilde{u}_{S}^{\gamma}(\mathbf{x},i)$ is the renormalized singlet 
feature function defined by
\begin{multline}
\label{eq:ubethe}
\tilde{u}_{S}^{\gamma}(\mathbf{x},i) =\\
\frac{1}{2}\sum_{\{j\}}\sum_{S'}\nu_{S,S'}^{\gamma}u_{S,S'}^{\gamma}(\mathbf{x},i,j) P(S' | S, \mathbf{x}, i, j).
\end{multline}
The conditional marginal probability $P(S' | S, \mathbf{x},i, j)$ 
is defined by
\begin{equation}
  \label{eq:condmarge}
  P(S' | S, \mathbf{x},i, j) = \frac{P(S, S'|\mathbf{x},i,j)}{P(S|\mathbf{x},i)}.
\end{equation}
Using $\tilde{u}_{S}^{\gamma}(\mathbf{x},i)$ and introducing a 
coupled external field $\tilde{\nu}_{S}^{\gamma}$, let us define a tentative 
total energy:
\begin{multline}
  \label{eq:energymf1}
  \tilde{E}(\mathbf{y},\mathbf{x},\theta) = -\sum_{\{i\}}\left[ 
 \sum_{\alpha}\lambda_{y_i}^{\alpha}s_{y_i}^{\alpha}(\mathbf{x},i)
 + \sum_{\gamma}\tilde{\nu}_{y_i}^{\gamma}\tilde{u}_{y_i}^{\gamma}(\mathbf{x}, i)
\right.\\
\left. + \sum_{\beta}\mu_{y_{i^{-}},y_i}^{\beta}t_{y_{i^{-}},y_i}^{\beta}(\mathbf{x}, i)
\right].
\end{multline}
By calculating the log-likelihood (Eq. \ref{eq:ll}) based on this energy and 
its derivative with respect to 
$\tilde{\nu}_{S}^{\gamma}$ (Eq. \ref{eq:dlambda}), 
and enforcing the optimality condition
$\partial L(\theta | \mathbf{x}, \mathbf{y})/\partial \tilde{\nu}_{S}^{\gamma} = 0$, we have
\begin{equation}
  \sum_{\{i\}}\tilde{u}_{S}^{\gamma}(\mathbf{x},i)\delta_{S,y_i} =
  \sum_{\{i\}}\tilde{u}_{S}^{\gamma}(\mathbf{x},i)P(S | \mathbf{x},i).
\end{equation}
Substituting this relation into Eq. (\ref{eq:bethe1}), we have
\begin{equation}
 \sum_{\{i<j\}}\sum_{\gamma}\nu_{y_i,y_j}^{\gamma}u_{y_i,y_j}^{\gamma}(\mathbf{x}, i, j)
 = 
\sum_{\{i\}}\sum_{\gamma}\tilde{u}_{y_i}^{\gamma}(\mathbf{x},i)
  \label{eq:bethe}
\end{equation}
Therefore, the pairwise energy terms can be converted into renormalized 
singlet energy terms as long as the alignment is optimal. 
For non-optimal alignments, we approximate the total energy by  
\begin{equation}
  \label{eq:eneapprox}
E(\mathbf{y},\mathbf{x},\theta) \approx \tilde{E}(\mathbf{y},\mathbf{x},\theta)
\end{equation}
with $\tilde{\nu}_{S}^{\gamma} = 1$.
The renormalized singlet feature function (Eq. \ref{eq:bethe}) explicitly 
accounts for the pairwise joint probability, and hence it may be called 
a Bethe or quasi-chemical approximation. 
Furthermore, if we assume two alignment sites are independent,
we can decouple the joint marginal probability as 
\begin{equation}
  \label{eq:decouple}
P(S, S'|\mathbf{x},i,j) \approx P(S|\mathbf{x},i)P(S'|\mathbf{x},j).
\end{equation}
This is a mean-field approximation.
Substituting this into Eqs. (\ref{eq:condmarge},\ref{eq:bethe}), we have
the following mean-field energy:
\begin{equation}
  \label{eq:umfa}
  \tilde{u}_{S}^{\gamma}(\mathbf{x},i) \approx
\frac{1}{2}\sum_{\{j\}}\sum_{S'}\nu_{S,S'}^{\gamma}u_{S,S'}^{\gamma}(\mathbf{x},i,j)
P(S' | \mathbf{x}, j).
\end{equation}
An advantage of this approximation is that we need not to compute the 
joint marginal probabilities.
By using either the Bethe (Eq. \ref{eq:ubethe}) or the mean-field 
(Eq. \ref{eq:umfa}) approximations, the energy of the alignment is expressed as 
\begin{multline}
  \label{eq:energymf}
  E(\mathbf{y},\mathbf{x},\theta) \approx -\sum_{\{i\}}\left[ 
 \sum_{\alpha}\lambda_{y_i}^{\alpha}s_{y_i}^{\alpha}(\mathbf{x},i)
 + \sum_{\gamma}\tilde{u}_{y_i}^{\gamma}(\mathbf{x}, i)\right.\\
\left. + \sum_{\beta}\mu_{y_{i^{-}},y_i}^{\beta}t_{y_{i^{-}},y_i}^{\beta}(\mathbf{x}, i)
\right].
\end{multline}
Note that there are apparently no external field parameters for the 
renormalized SFFs ($\tilde{u}_S^{\gamma}(\cdot)$); they are included in 
the definitions (Eqs. \ref{eq:ubethe}, \ref{eq:umfa}).
Since the mean-field feature functions are effectively singlet feature 
functions, we can apply the standard procedure for learning and alignment, 
provided that the mean-fields are known. Of course, the mean-fields are 
not known in advance so that we need to obtain the partition function
by an iterative procedure. That is, 
\begin{enumerate}
\item Arbitrarily set $\tilde{u}_S^\gamma(\cdot)$.
\item Calculate the partition function and marginal probabilities based on 
the previously calculated $\tilde{u}_S^\gamma(\cdot)$.
\item Based on the partition function and marginal probabilities in the 
previous step, update $\tilde{u}_S^{\gamma}(\cdot)$ by Eq. (\ref{eq:ubethe}) 
or Eq. (\ref{eq:umfa}).
\item Iterate steps 2 and 3 until convergence.
\end{enumerate}
The algorithms for computing the partition function and marginal 
probabilities are a subject of the next section.

\section*{Algorithms for alignment and learning}
\subsection*{Computation of partition function, marginal probabilities and optimal alignment}
The partition function (Eq. \ref{eq:pfunc}) and marginal probabilities can be
calculated efficiently by dynamic programming (or transfer matrix method).
In this section, we assume that pairwise terms are approximated as 
renormalized SFFs (Eqs. \ref{eq:ubethe}, \ref{eq:umfa}), and they are treated 
as ordinary SFFs.
First we define the transfer matrix:
\begin{equation}
  \label{eq:tmat}
  T_{i}(S,S') = \exp [ e_{i}(S,S') / T ]
\end{equation}
where $S',S \in \mathcal{S}$, $T$ is the temperature, and
\begin{equation}
  \label{eq:texponent}
e_{i}(S,S') = \sum_{\alpha}\lambda_{S'}^{\alpha} s_{S'}^{\alpha}(\mathbf{x},i) +
\sum_{\beta}\mu_{S,S'}^{\beta}t_{S,S'}^{\beta}(\mathbf{x},i).
\end{equation}
The partition function (Eq. \ref{eq:pfunc}) is then expressed as 
\begin{equation}
\label{eq:pfunct}
  Z(\mathbf{x}) = \sum_{\{\mathbf{y}\}}\prod_{\{i\}}T_i(y_{i^{-}},y_{i})
\end{equation}
where the summation is over all possible model states of each residue of the 
target sequence.
In order to compute the partition function Eq. (\ref{eq:pfunct}), we define
an auxiliary function $Z_{i,j}(S_k, S_l)$ where $i,j = 0,\cdots, L+1$ and 
$S_k\in \{M_k, I_k, D_k\}$, $S_l \in \{M_l, I_l, D_l\}$. 
$Z_{i,j}(S_k, S_l)$ is the partition function of the 
sub-sequence $x_ix_{i+1}\cdots x_{j}$ where its termini $i$ and $j$ 
are fixed to the model states $S_k$ and $S_l$, respectively. These 
conditions are given as 
\begin{eqnarray}
  Z_{i,i}(S_k,S) & = & \delta_{S_k,S} \label{eq:bc_faux},\\
  Z_{j,j}(S,S_l) & = & \delta_{S,S_l} \label{eq:bc_baux}.
\end{eqnarray}
By the construction of the model, the following boundary conditions hold 
in particular:
\begin{eqnarray}
Z_{0,0}(M_0, M_0) & = &1,\\
Z_{L+1,L+1}(M_{M+1}, M_{M+1}) & = &1.
\end{eqnarray}
The partition function $Z(\mathbf{x})$ is given as 
\begin{equation}
  Z(\mathbf{x}) = Z_{0,L+1}(M_0, M_{M+1}).
\end{equation}
Based on the boundary condition Eq. (\ref{eq:bc_faux}),
the following forward recurrence equations for $Z_{i,j}(S_k,M_{l})$ hold for
$j = i, \cdots, L+1$ and $l=k,\cdots, M+1$:
\begin{multline}
  Z_{i,j}(S_k, M_{l}) = Z_{i,j-1}(S_k, M_{l-1}) T_{j}(M_{l-1},M_{l})\\
 + Z_{i,j-1}(S_k, I_{l-1}) T_{j}(I_{l-1},M_{l}) \\
 + Z_{i,j-1}(S_k, D_{l-1}) T_{j}(D_{l-1},M_{l}); \label{eq:faux1}
\end{multline}
\begin{multline}
  Z_{i,j}(S_k, I_{l})  = Z_{i,j-1}(S_k, M_{l}) T_{j}(M_{l},I_{l}) \\
+ Z_{i,j-1}(S_k, I_{l}) T_{j}(I_{l},I_{l}) \\
+ Z_{i,j-1}(S_k, D_{l}) T_{j}(D_{l},I_{l}); \label{eq:faux2}
\end{multline}
\begin{multline}
  Z_{i,j}(S_k, D_{l}) =  Z_{i,j}(S_k, M_{l-1}) T_{j}(M_{l-1},D_{l}) \\
+ Z_{i,j}(S_k, I_{l-1}) T_{j}(I_{l-1},D_{l}) \\
+ Z_{i,j}(S_k,D_{l-1}) T_{j}(D_{l-1},D_{l}). \label{eq:faux3}
\end{multline}
It is understood that the terms involving non-existent states
and/or incompatible state transitions
(e.g, $Z_{1,1}(M_0, D_0), Z_{1,0}(I_0, I_0)$, etc.) are ignored.
Similarly, together with the boundary condition Eq. (\ref{eq:bc_baux}), 
the backward recurrence equations are given for $i= j, \cdots, 0$ and 
$k = l,\cdots,0$ as
\begin{multline}
  Z_{i,j}(M_k, S_{l}) 
   =  T_{i+1}(M_{k},M_{k+1})Z_{i + 1,j}(M_{k+1}, S_{l})  \\
   +  T_{i+1}(M_{k},I_{k})Z_{i + 1,j}(I_{k}, S_{l})  \\
   +  T_{i}(M_{k},D_{k+1})Z_{i,j}(D_{k+1}, S_{l});  \label{eq:baux1}
\end{multline}
\begin{multline}
  Z_{i,j}(I_k, S_{l}) 
   = T_{i+1}(I_{k},M_{k+1})Z_{i + 1,j}(M_{k+1}, S_{l})\\
   + T_{i+1}(I_{k},I_{k})Z_{i + 1,j}(I_{k}, S_{l})  \\
   + T_{i}(I_{k},D_{k+1})Z_{i,j}(D_{k+1}, S_{l});  \label{eq:baux2}
\end{multline}
\begin{multline}
  Z_{i,j}(D_k,S_{l}) 
   =  T_{i+1}(D_{k},M_{k+1})Z_{i + 1,j}(M_{k+1}, S_{l})  \\
   +  T_{i+1}(D_{k},I_{k})Z_{i + 1,j}(I_{k}, S_{l})  \\
   +  T_{i}(D_{k},D_{k+1})Z_{i,j}(D_{k+1}, S_{l}).  \label{eq:baux3}
\end{multline}
For convenience, let us define the forward auxiliary function $F_i(S_k)$ 
and the backward auxiliary function $B_i(S_k)$ by
\begin{eqnarray}
  F_i(S_k) & = & Z_{0,i}(M_0, S_k),  \label{eq:faux}\\
  B_i(S_k) & = & Z_{i,L+1}(S_k, M_{M+1}).\label{eq:baux}
\end{eqnarray}
Using $F_i$ and $B_i$, and $Z_{i,j}$, we can calculate marginal probabilities.
The joint marginal probability is obtained as
\begin{equation}
  \label{eq:jointprob}
P(S_k, S_l|\mathbf{x},i,j) = \frac{%
F_i(S_k)Z_{i,j}(S_k, S_l) B_j(S_l)}{Z(\mathbf{x})}.
\end{equation}
In particular, when $i=j$ and $S_k = S_l$, we have 
\begin{equation}
  P(S_k | \mathbf{x},i) = \frac{F_i(S_k)B_i(S_k)}{Z(\mathbf{x})}.
\end{equation}
Similarly, for states with allowed transitions $S$ and $S'$
(Table \ref{tab:transition}),
\begin{equation}
  P(S, S' | \mathbf{x},i) =
\frac{F_{i^{-}}(S)T_{i}(S,S')B_i(S')}{Z(\mathbf{x})}.\label{eq:djprob}
\end{equation}
Using these marginal probabilities, the renormalized SFFs for pairwise
terms (Eqs. \ref{eq:ubethe}, \ref{eq:umfa}) can be computed.

The optimal alignment for a given model and a target sequence is 
the one that yields the minimum energy, which corresponds to the free energy
of the system at zero temperature ($T = 0$).
The recurrence equations for the optimal alignment can be derived as 
the zero-temperature limit of the forward 
recurrence equations using the following formula\cite{HirotaANDTakahashi}: 
\begin{equation}
  \label{eq:limmax}
  \lim_{\epsilon \to +0}\epsilon\log[\sum_{i}e^{a_i/\epsilon}] = \max_{i}a_i.
\end{equation}
That is, if we define a function 
\begin{equation}
  \label{eq:optene}
A_{i}(S_k) = \lim_{T\to 0}[T \log F_{i}(S_k)],
\end{equation}
the energy of the optimal alignment 
$\mathcal{A} = (\mathbf{x},\mathbf{y}_{\mathrm{opt}})$ is given by
\begin{equation}
  \label{eq:optfinal}
E(\mathbf{y}_{\mathrm{opt}}, \mathbf{x}) = -A_{L+1}(M_{M+1}).
\end{equation}
More concretely, we first set the boundary condition
\begin{eqnarray}
  \label{eq:optbc}
  A_0(M_0) & = & 0,
\end{eqnarray}
and apply the zero-temperature limit to the both sides of 
the forward recurrence equations for
$F_{i}(S_k) = Z_{0,i}(M_0,S_k)$ 
(Eqs. \ref{eq:faux1}--\ref{eq:faux3}),  we have
\begin{multline}
  \label{eq:optali}
  A_i(M_k) = \max \{
      [A_{i-1}(M_{k-1}) + e_{i}(M_{k-1},M_{k})],\\
      [A_{i-1}(I_{k-1}) + e_{i}(I_{k-1},M_{k})],\\
      [A_{i-1}(I_{k-1}) + e_{i}(D_{k-1},M_{k})]\};
\end{multline}
\begin{multline}
  A_i(I_k) =  \max \{
      [A_{i-1}(M_k) + e_{i}(M_{k},I_{k})],\\
      [A_{i-1}(I_k) + e_{i}(I_{k},I_{k})],\\
      [A_{i-1}(D_k) + e_{i}(D_{k},I_{k})]\};
\end{multline}
\begin{multline}
  A_{i}(D_k) = \max \{
      [A_{i}(M_{k-1}) + e_{i}(M_{k-1},D_{k})],\\
      [A_{i}(I_{k-1}) + e_{i}(I_{k-1},D_{k})],\\
      [A_{i}(D_{k-1}) + e_{i}(D_{k-1},D_{k})]\}.
\end{multline}
By tracing back the site-state pairs that yield the optimal values of 
$A_{i}(S_k)$ at each step, we can find the optimal alignment 
$\mathbf{y}_{\mathrm{opt}}$.

\subsection*{Parameter learning with multiple sequence alignment}
\paragraph{Global optimization of parameters.}
The parameters of a profile CRF are the set of  external fields 
$\lambda_{S}^{\alpha}$, $\mu_{S',S}^{\beta}$ and $\nu_{S,S'}^{\gamma}$
(of course, we need to specify the feature functions to start to with). 
The input for parameter learning is a multiple sequence alignment (MSA)
of a protein family, from which the model architecture 
 must be somehow specified ``by hand.''
In this process, we need to specify which columns of the MSA correspond 
to matching states. After the columns of matching states are determined, 
matching, insertion and deletion states can be assigned to each
column of each sequence in the MSA.

After the model architecture has been determined, the learning can be done by 
maximizing the likelihood using the sequences of the input MSA.
Let these alignments be $(\mathbf{x}^{(p)},\mathbf{y}^{(p)})$ where 
$p = 1,\cdots, n$ is the index of sequences. 
The joint log-likelihood is given by
\begin{multline}
  \label{eq:lltot}
  L(\theta|\{\mathbf{x}^{(p)},\mathbf{y}^{(p)}\}) = \\
-\sum_{k=1}^{n}\left[E(\mathbf{y}^{(p)},\mathbf{x}^{(p)},\theta) 
+ \log Z(\mathbf{x}^{(p)},\theta)\right].
\end{multline}
Since the total energy is a linear function of the parameters, and 
$-\log Z$ is the free energy of the system which is always convex, 
the total log-likelihood is also a convex function of the parameters.
This implies that we can obtain the globally optimal parameter sets by 
gradient-based methods. In practice, minimizing the bare log-likelihood may 
results in over-fitting of the parameters to the training set. Therefore, 
we define an alternative objective function $K(\theta|\{\mathbf{x}^{(p)},\mathbf{y}^{(p)}\})$ which includes prior probability density of 
the parameters for regularization:
\begin{equation}
  \label{eq:llbayes}
  K(\theta|\{\mathbf{x}^{(p)},\mathbf{y}^{(p)}\}) =
L(\theta|\{\mathbf{x}^{(p)},\mathbf{y}^{(p)}\}) + \log P(\theta)
\end{equation}
where $P(\theta)$ is a Gaussian prior:
\begin{multline}
  \label{eq:prior}
P(\theta) = 
  \prod_{\alpha,S}\exp\left[-\frac{(\lambda_{S}^{\alpha})^2}{2(\sigma_{S}^{\alpha})^2}\right]
\prod_{\beta,S',S}\exp\left[-\frac{(\mu_{S',S}^{\beta})^2}{2(\sigma_{S',S}^{\beta})^2}\right]\\
\times
\prod_{\gamma,S,S'}\exp\left[-\frac{(\nu_{S,S'}^{\beta})^2}{2(\sigma_{S,S'}^{\gamma})^2}\right].
\end{multline}
Here, the hyper-parameters $\sigma_{S}^{\alpha}$, $\sigma_{S',S}^{\beta}$ 
and $\sigma_{S,S'}^{\gamma}$
are the (expected) standard deviations of the corresponding external fields, 
and must be specified \emph{a priori} (however, if we use a hierarchical Bayes 
model, these hyper-parameters can be automatically adjusted based on 
the training data).
Since we can calculate the gradient of this log-likelihood, it is possible to
use gradient-based optimization techniques. 
Since $K(\theta|\{\mathbf{x}^{(p)},\mathbf{y}^{(p)}\})$ (Eq. \ref{eq:llbayes}) 
is still convex, the globally optimal parameters are guaranteed to be 
found by gradient descent methods.

\paragraph{Bayesian learning.}
It is also possible to apply the Bayesian learning framework\cite{Neal1996}. 
That is, instead of using a single, globally optimal parameter set, 
we can use a set of suboptimal parameters to make robust predictions.
From Bayes' formula, we have
\begin{equation}
  P(\theta|\{\mathbf{x}^{(p)},\mathbf{y}^{(p)}\}) \propto
 \exp[L(\theta|\{\mathbf{x}^{(p)},\mathbf{y}^{(p)}\})] P(\theta).
\end{equation}
Using this equation, a Bayesian alignment for the target sequence $\mathbf{x}$
 may be selected so as to maximize the following probability:
\begin{equation}
  P(\mathbf{y}|\mathbf{x},\{\mathbf{x}^{(p)},\mathbf{y}^{(p)}\}) 
= \int\! P(\mathbf{y}|\mathbf{x},\theta)P(\theta|\{\mathbf{x}^{(p)},\mathbf{y}^{(p)}\})\mathrm{d}\theta.
\end{equation}
Suboptimal parameters may be obtained by Markov chain Monte Carlo simulations
in the $\theta$-space, 
using $-\log K(\theta|\{\mathbf{x}^{(p)},\mathbf{y}^{(p)}\})$ as 
the ``energy'' of the system. 
Since the gradients of the log-likelihood can be computed, a hybrid Monte Carlo
method is also at our disposal for efficient sampling.

We can also employ hierarchical Bayes learning which can automatically 
adjust the the hyper-parameters for the prior,
$\sigma_{S}^{\alpha}$ and $\sigma_{S,S'}^{\beta}$, based on the training 
set\cite{Neal1996}.

\section*{Discussion}
In this paper, we have formulated the profile CRF to model protein 
families with possible long-range correlations such as structural information.
The profile CRF model is clearly an extension of both the molecular field
theory of Finkelstein and Reva (FR theory)\cite{FinkelsteinANDReva1991,FinkelsteinANDReva1996a,FinkelsteinANDReva1996b} and the profile HMM\cite{KroghETAL1994,DurbinETAL},
and hence an integration of these. Here, we shall discuss the relationship 
of the present model with these two earlier models.

The FR theory is particularly focused on 3D structures of proteins.
Accordingly, its model is explicitly represented in the 3D space as a set 
of lattice points. The lattice points mostly correspond to residues in 
secondary structure elements (SSEs), and these points may be regarded as 
``match'' states in the present framework. The FR model does not allow
gaps within each SSE, only insertions are allowed in the regions between 
two SSEs. The energy functions ($\approx$ feature functions) are 
physics-based ones, and the parameters are not optimized 
to fit some training data, but obtained from physical experiments.
Therefore, the FR models are more suitable for studying physical aspects 
of protein folding and structure prediction, but less so for more 
general-purpose sequence analysis. Nevertheless, almost all the theoretical
foundations of the FR theory such as calculation of partition functions, 
marginal probabilities, mean-field approximations, 
but except for parameter learning, are shared by profile CRFs.
After all, the both models are extensions of the 1D Ising model.

The analogy between 1D Ising model and a more general sequence alignment 
problem was pointed out by Miyazawa\cite{Miyazawa1995}, which was further 
extended to the problem of sequence-structure alignment with a mean-field 
approximation\cite{MiyazawaANDJernigan2000}.
Later, Koike \emph{et al.}\cite{KoikeETAL2004} applied this analogy to 
compute partition functions and marginal probabilities in protein structure 
comparison with the Bethe approximation. 
By complementing the FR theory with these techniques, the alignment 
algorithm can be made more general, and one such generalization is the 
profile CRF model.
The improvements made by profile CRFs on the FR theory are thus clear: 
more general treatment of model states, possible insertions and deletions 
at any sites, and parameter learning based on MSA.

Profile HMMs, being a class of generative models, need to calculate the joint
probability of alignment $P(\mathbf{x},\mathbf{y})$ while profile CRFs, being 
a class of discriminative model, directly calculates the conditional 
probability $P(\mathbf{y} | \mathbf{x})$.
In special cases, with the definition of the conditional probability 
$P(\mathbf{y} | \mathbf{x}) = P(\mathbf{x},\mathbf{y})/P(\mathbf{x})$ in mind, 
we may regard $Z(\mathbf{x})$ as $P(\mathbf{x})$ and 
$\exp[-E(\mathbf{y},\mathbf{x})]$ as $P(\mathbf{x},\mathbf{y})$.
More specifically, if we define only the following feature functions 
(and no others) with appropriate values for external fields,
we can construct a CRF that is equivalent to a given HMM:
\begin{enumerate}
\item Define singlet feature functions $s_{S_k}^{x}$ 
for matching and insertion states as in Eq. (\ref{eq:singlet1}). 
For deletion states, just define a constant SFF (always equal to 1).
\item Define sequence-independent feature functions $t_{S_k,S_l}^{-}$ 
for each transitions as in Eq. (\ref{eq:doublet1}).
\item Set the singlet external fields as $\lambda_{S_k}^{R} = \log q_{S_k}(R)$ 
($q_{S_k}(R)$: the emission probability of the HMM).
\item Set the doublet external fields as $\mu_{S_k,S_l}^{-} = \log p_{S_k,S_l}$ 
($p_{S_k,S_l}$: transition probability of the HMM).
\end{enumerate}
However, this equivalence breaks down as soon as we incorporate other feature
functions into profile CRFs since the Boltzmann factor 
$\exp[-E(\mathbf{y},\mathbf{x})]$ may no longer satisfy a condition of 
probability measure (i.e., normalization to 1). 
Thus, HMMs are a very special class of CRFs.

In summary, we have presented the profile CRF model. This model is flexible
enough to accommodate almost any features of target sequences including PSSM,
local sequence patterns, and even long-range correlations. 
It can also incorporate various features of a modeled protein family such as 
local structures and long-range pairwise interactions. Although concrete 
implementations are yet to be done, we expect this model to be a useful 
alternative to conventional methods for analyzing and understanding 
protein sequences and structures.

\section*{Acknowledgments}
The author thanks Drs. Takeshi Kawabata, Ryotaro Koike, Kengo Kinoshita and 
Motonori Ota for their valuable comments on the first draft of this manuscript.


\begin{thebibliography}{10}

\bibitem{NeedlemanANDWunsch1970}
Needleman, S.~B. and Wunsch, C.~D.
\newblock A general method applicable to the search for similarities in the
  amino acid sequence of two proteins.
\newblock {\em J. Mol. Biol.}{ \bf 48}, 443--453, 1970.

\bibitem{SmithANDWaterman1981}
Smith, T.~F. and Waterman, M.~S.
\newblock Identification of common molecular subsequences.
\newblock {\em J. Mol. Biol.}{ \bf 147}, 195--197, 1981.

\bibitem{Gotoh1982}
Gotoh, O.
\newblock An improved algorithm for matching biological sequences.
\newblock {\em J. Mol. Biol.}{ \bf 162}, 705--708, 1982.

\bibitem{DurbinETAL}
Durbin, R., Eddy, R., Krogh, A., and Mitchison, G.
\newblock Biological Sequence Analysis: Probabilistic Models of Proteins and
  Nucleic Acids.
\newblock Cambridge Univ. Press, Cambridge, U. K., , 1999.

\bibitem{ProteinBioinfo}
Eidhammer, I., Jonassen, I., and Taylor, W.~R.
\newblock Protein bioinformatics.
\newblock Wiley \& Sons, Chichester, England, , 2004.

\bibitem{GribskovETAL1987}
Gribskov, M., Mc{L}achlan, A.~D., and Eisenberg, D.
\newblock Profile analysis: Detection of distantly related proteins.
\newblock {\em Proc. Natl. Acad. Sci. U.S.A.}{ \bf 84}, 4355--4358, 1987.

\bibitem{AltschulETAL1997}
Altschul, S.~F., Madden, T.~L., Schaffer, A.~A., Zhang, J., Zhang, Z., Miller,
  W., and Lipman, D.~L.
\newblock Gapped blast and {PSI}-blast: A new generation of protein database
  search programs.
\newblock {\em Nucleic Acids Res.}{ \bf 25}, 3389--3402, 1997.

\bibitem{KinjoANDNakamura2008}
Kinjo, A.~R. and Nakamura, H.
\newblock Nature of protein family signatures: Insights from singular value
  analysis of position-specific scoring matrices.
\newblock {\em PLoS ONE}{ \bf 3}, e1963, 2008.

\bibitem{KroghETAL1994}
Krogh, A., Brown, M., Mian, I.~S., Sj\"olander, K., and Haussler, D.
\newblock Hidden {Markov} models in computational biology. {A}pplications to
  protein modeling.
\newblock {\em J. Mol. Biol.}{ \bf 235}, 1501--1531, 1994.

\bibitem{BowieETAL1991}
Bowie, J.~U., L{\"{u}}thy, R., and Eisenberg, D.
\newblock A method to identify protein sequences that fold into a known
  three-dimensional structure.
\newblock {\em Science}{ \bf 253}, 164--170, 1991.

\bibitem{JonesETAL1992}
Jones, D.~T., Taylor, W.~R., and Thornton, J.~M.
\newblock A new approach to protein fold recognition.
\newblock {\em Nature}{ \bf 358}, 86--89, 1992.

\bibitem{CRF}
Lafferty, J., {McCallum}, A., and Pereira, F.
\newblock Conditional random fields: Probabilistic models for segmenting and
  labeling sequence data.
\newblock In {\em Proc. Int. Conf. Machine Learning},  282--289, , 2001.

\bibitem{DoETAL2006}
Do, C., Gross, S., and Batzoglou, S.
\newblock {CONTRAlign}: Discriminative training for protein sequence alignment.
\newblock In {\em Proceedings of the Tenth Annual International Conference on
  Computational Molecular Biology (RECOMB 2006)},  160--174, , 2006.

\bibitem{DeCaprioETAL2007}
{DeCaprio}, D., Vinson, J.~P., Pearson, M.~D., Montgomery, P., Doherty, M., and
  Galagan, J.~E.
\newblock Conrad: Gene prediction using conditional random fields.
\newblock {\em Genome Res.}{ \bf 17}, 1389--1398, 2007.

\bibitem{ZhaoETAL2008}
Zhao, F., Li, S., Sterner, B.~W., and Xu, J.
\newblock Discriminative learning for protein conformation sampling.
\newblock {\em Proteins}{ \bf 73}, 228--240, 2008.

\bibitem{FinkelsteinANDReva1991}
Finkelstein, A.~V. and Reva, B.~A.
\newblock A search for the most stable folds of protein chains.
\newblock {\em Nature}{ \bf 351}, 497--499, 1991.

\bibitem{FinkelsteinANDReva1996a}
Finkelstein, A.~V. and Reva, B.~A.
\newblock A search for the most stable folds of protein chains: {I}.
  application of a self-consistent molecular field theory to a problem of
  protein three-dimensional structure prediction.
\newblock {\em Protein Eng.}{ \bf 9}, 387--397, 1996.

\bibitem{FinkelsteinANDReva1996b}
Finkelstein, A.~V. and Reva, B.~A.
\newblock A search for the most stable folds of protein chains: {II}.
  computation of stable architectures of beta-proteins using a self-consistent
  molecular field theory.
\newblock {\em Protein Eng.}{ \bf 9}, 399--411, 1996.

\bibitem{KawabataANDDoi1997}
Kawabata, T. and Doi, J.
\newblock Improvement of protein secondary structure prediction using binary
  word encoding.
\newblock {\em Proteins}{ \bf 27}, 36--46, 1997.

\bibitem{HirotaANDTakahashi}
Hirota, R. and Takahashi, D.
\newblock Discrete and Ultradiscrete Systems.
\newblock Kyoritsu Shuppan Co., Tokyo, Japan, , 2003.
\newblock In Japanese.

\bibitem{Neal1996}
Neal, R.~M.
\newblock Bayesian Learning for Neural Networks.
\newblock Number 118 in Lecture Notes in Statistics. Springer-Verlag, New York,
  U.S.A., , 1996.

\bibitem{Miyazawa1995}
Miyazawa, S.
\newblock A reliable sequence alignment method based on probabilities of
  residue correspondences.
\newblock {\em Protein Eng.}{ \bf 8}, 999--1109, 1995.

\bibitem{MiyazawaANDJernigan2000}
Miyazawa, S. and Jernigan, R.~L.
\newblock Identifying sequence-structure pairs undetected by sequence
  alignments.
\newblock {\em Protein Eng.}{ \bf 13}, 459--475, 2000.

\bibitem{KoikeETAL2004}
Koike, R., Kinoshita, K., and Kidera, A.
\newblock Probabilistic description of protein alignments for sequences and
  structures.
\newblock {\em Proteins}{ \bf 56}, 157--166, 2004.

\end{thebibliography}

\end{document}